\documentclass[preprint,showpacs,preprintnumbers,amsmath,amssymb,superscriptaddress,nofootinbib]{revtex4-1}
\pdfoutput=1
\usepackage{graphicx,subfigure}
\usepackage{dcolumn}
\usepackage{bm}
\usepackage{slashed}
\usepackage{color}

\newcommand{\be}{\begin{eqnarray}}
\newcommand{\ee}{\end{eqnarray}}
\newcommand{\bea}{\begin{eqnarray}}
\newcommand{\eea}{\end{eqnarray}}

\newcommand{\MeV}{{~\rm MeV}}

\newcommand{\gev}{{~\rm GeV}}
\newcommand{\tev}{{~\rm TeV}}

\newcommand{\lsim}{\begin{array}{c}\,\sim\vspace{-26pt}\\< \end{array}}
\newcommand{\gsim}{\begin{array}{c}\sim\vspace{-26pt}\\> \end{array}}

\newcommand{\ias}{School of Natural Sciences, Institute for Advanced Study, Princeton, NJ 08540, USA}
\newcommand{\nyu}{CCPP, Department of Physics, New York University, New York, NY 10003, USA}
\newcommand{\jhu}{Dept.~of Physics \& Astronomy, Johns Hopkins University, Baltimore, MD  21218, USA}

\begin{document}

\title{Simply Unnatural Supersymmetry}
\author{Nima Arkani-Hamed}
\email{arkani@ias.edu}
\affiliation{\ias}
\author{Arpit Gupta}
\email{arpit@pha.jhu.edu}
\affiliation{\jhu}
\author{David E. Kaplan}
\email{dkaplan@pha.jhu.edu}
\affiliation{\jhu}
\author{Neal Weiner}
\email{weiner@physics.nyu.edu}
\affiliation{\nyu}
\author{Tom Zorawski}
\email{tz137@pha.jhu.edu}
\affiliation{\jhu}

\begin{abstract}
The current measurement of the Higgs mass, the ubiquitous nature of loop-suppressed gaugino masses in gravity-mediated supersymmetry breaking,  relic dark matter density from $\sim$ TeV mass gauginos, together with the success of supersymmetric gauge coupling unification, suggest that scalar superpartner masses are roughly $m_{sc} \sim$ 100-1000 TeV.  Higgsino masses, if not at the Planck scale, should generically appear at the same scale. The gaugino mass contributions from anomaly mediation, with the heavy Higgsino threshold, generally leads to a more compressed spectrum than standard anomaly mediation, while the presence of extra vector-like matter near $m_{sc}$ typically leads to an even more compressed spectrum.  Heavy Higgsinos improve gauge coupling unification relative to the MSSM. Heavy scalars suggest new possibilities for flavor physics -- large flavor violations in the soft masses are now allowed, yielding interesting levels for new FCNC's, and re-opening the attractive possibility of a radiatively generated fermion mass hierarchy.  Gluinos and binos/winos must decay through higher dimension operators, giving a unique handle on the presence of new physics the scale $m_{sc}$. Gluino decays should be spectacular, for example yielding events with four tops -- at modestly displaced vertices -- and two Higgses plus missing energy.  The high scale $m_{sc}$ can also be probed in flavor-violating gluino decays, as well as a specific pattern of gluino branching ratios to the third generation. Finally, with heavy Higgsinos, the dominant decay for neutral winos and binos proceeds via the Higgs $\tilde b \to \tilde w h$. The heaviness of the Higgsinos can be inferred from the branching ratio for the rare decay $\tilde b \to \tilde w Z$.
\end{abstract}

\maketitle

\section{Introduction}

The central drama of physics at the TeV scale is the question of naturalness. Is the mass of the Higgs the result of an accidental tuning of parameters in the far ultraviolet, or due to dynamics beyond the standard model at the weak scale?  A preponderance of evidence -- the lack of significant hints for physics beyond the standard model in electroweak precision tests, studies of lepton and quark flavor physics, measurements of CP violation, and searches for new states at the LHC and other colliders -- suggests that  electroweak sector may well be fine-tuned to some extent.  The incredibly small value of the cosmological constant is a much more dramatic challenge to our notion of naturalness. The best explanation to date for minuscule size of the cosmological constant  is anthropically motivated fine-tuning \cite{Weinberg:1987dv} as a vacuum selection mechanism in an enormous landscape of vacua \cite{Bousso:2000xa, Susskind:2003kw}.  This, together with the existence of reasonable anthropic selection effects on the value of the electroweak scale \cite{Agrawal:1997gf}, adds to the plausibility of fine-tuning at the TeV scale. But how much fine-tuning should we expect? What is the actual UV cutoff for quadratically divergent contributions to the Higgs mass?                         
                                                                                                                                                                
The original motivation for the minimal supersymmetric standard model (MSSM) \cite{Dimopoulos:1981zb} was to solve the Higgs naturalness problem with superpartners at the weak scale. With it came a compelling candidate for dark matter, and a prediction for the weak mixing angle from gauge coupling unification that was borne out by LEP.  However, the model dependence of the dark-matter abundance, and the logarithmic dependence of unification on mass scales mean that these concrete successes of the MSSM are not tightly tied to a natural superpartner spectrum.  In fact, the Higgs mass $m_H \sim$ 125 GeV already requires some tuning in the MSSM, or some significant departure from it.  

On the other hand, $m_H \sim 125$ GeV falls into the range of Higgs masses predicted by the `split supersymmetry' versions of the MSSM (\cite{Wells:2003tf,ArkaniHamed:2004fb,Giudice:2004tc}), which have spectra with fine-tuning for electroweak symmetry breaking,  but which preserve the gauge-coupling unification and dark matter predictions of the model.                                                                                                                                                          
                                                                                                                                                                
Of course, completely natural supersymmetric theories may still turn out to describe physics at the TeV scale, and there have been no shortage of models of this sort proposed recently in response to null-results for new physics from the LHC. It is however fair to say that these models are rather elaborate. Many of these theories are actually just as fine-tuned as more conventional versions of supersymmetry, but the tuning is more hidden.  The more sensible theories of this sort may be ``natural" with respect to variations of their Lagrangian parameters, but in an admittedly hard-to-quantify sense, their epicyclic character involves a tuning in ``model space." By contrast, with split supersymmetry, we take the minimal field content of the supersymmetric standard model with no additional decorations or dynamical mechanisms. There is no tuning in ``model space". The theory is of course unapologetically and explicitly finely tuned for electroweak symmetry breaking, but this tuning has a clear anthropic purpose.                                                                                                                  
                                                                                                                                                                
In this paper, we describe and explore the simplest picture of the the world arising from fine-tuned supersymmetric theories. Our guiding principle is that the model should be ``simply un-natural". There is an explicit, un-natural tuning for the weak scale with a clear ``environmental" purpose, but in every other way the theoretical structure should be as simple as possible.  To this end we will  follow where the theory leads us, without any clever model-building gymnastics. Following what theories of supersymmetry breaking ``want to do"  leads us to theories with a ``minimally split" spectrum where gauginos are near 1 TeV, while scalars and Higgsino and gravitino parametrically heavier by a loop factor, at a scale $m_{sc}$ between $\sim 10^2-10^3$ TeV. This kind of spectrum has long been a ubiquitous feature of simple, concrete models of SUSY breaking.  Its modern manifestation was in the context of theories with anomaly mediated SUSY breaking \cite{Giudice:1998xp}, {\it without} the clever sequestering mechanism of \cite{Randall:1998uk}.

In \cite{Giudice:1998xp}, the heavy scalars were thought of as something of an embarrassment. This spectrum was later proposed as a serious possibility for supersymmetric theories in \cite{Wells:2003tf,ArkaniHamed:2004fb,Giudice:2004tc}, put forward as the ``simplest model of split SUSY" in \cite{ArkaniHamed:2006mb}, and further studied in \cite{Kaplan:2006vm}. We re-initiated a study of this scenario in \cite{Madrid,SavasFest}. For obvious reasons, this spectrum has received renewed attention of late \cite{Acharya:2008bk,Hall:2011jd,Kane:2011kj,Bhattacherjee:2012ed,Arvanitaki:2012ps,Hall:2012zp}. The Higgs mass prefers this ``minimally split" spectrum, rather than the more radical possibility  scalars up to around $\sim 10^{13}$ GeV \cite{ArkaniHamed:2004fb}. This is perfectly in line with the ``simply un-natural" perspective, since theories with much heavier scalars needed extra theoretical structure to suppress gaugino masses by much more than a loop factor relative to the gravitino mass. 

With this split spectrum, gaugino masses receive comparable contributions from  anomaly mediation and the heavy Higgsinos, as well as other possible vector-like matter near the scale $m_{sc}$. As we will see, this picture has important consequences for flavor physics, as well as a host of novel collider signals that constrain the scale $m_{sc}$ in an interesting way.

\section{Simplest Tuned Picture of the World}

\subsection{Model and Spectrum}


Supersymmetry breaking must give all superpartners in the MSSM masses above their current bounds.  Once supersymmetry is broken, there are no symmetries protecting the sfermion masses, and thus scalar masses are expected at some level.  On the other hand, (Majorana) gaugino masses require the breaking of an R symmetry, and are thus not guaranteed to arise at the same level.   In the case where supersymmetry breaking is communicated via irrelevant operators suppressed by a scale $M$, sfermion and gaugino masses could arise from the operators of the form
\begin{equation}\label{gravmed}
\int d^4 \theta \frac{X^\dagger X Q^\dagger Q}{M^2}  \:\:\:\:\:  {\rm and } \:\:\:\:\: \int d^2\theta \frac{Y W^\alpha W_\alpha}{M},
\end{equation}
where $Q$ and $W^\alpha$ represent visible sector matter and gauge superfields respectively and $X$ and $Y$ are hidden sector chiral superfields which have non-zero vevs in their auxiliary (`$F$') components.  There are no requirements (other than the absence of a shift symmetry) on the quantum numbers of $X$, and thus it could be any field from the hidden sector.  On the other hand, $Y$ is required to be an exact gauge and global singlet.  This stringent requirement makes it clear that gaugino masses will typically not be the same size as scalar masses for generic  hidden sectors.  In fact, most models of supersymmetry breaking sectors do not contain such singlets \cite{Affleck:1984xz,Affleck:1983rr, Affleck:1983mk, Affleck:1983vc,Affleck:1984uz,Affleck:1984mf,Intriligator:2006dd,Intriligator:2007py}, and this affects both gravity and gauge mediation.  While this problem has been `solved', in the sense that models generating larger gaugino masses have been found \cite{Nelson:1995ar,Intriligator:1996pu}--with non-generic superpotentials and/or many discrete symmetries imposed--we take the position that generic models of supersymmetry breaking produce much larger scalar masses than gaugino masses, that is,  this is what the models {\it want} to do.  In line with our ``simply un-natural" philosophy, we assume the theory-space tuning required to have degenerate sfermions and gauginos is more severe than that required to get the correct electroweak scale.

Another contribution to superpartner masses that theories of broken supersymmetry ``want to generate" arises from anomaly mediation. The breaking of $R$ symmetry associated with tuning away the cosmological constant with a constant superpotential gives rise to gaugino masses  of order a loop factor beneath the gravitino mass.  While there are clever ways to suppress this contribution \cite{Luty:2002ff,ArkaniHamed:2004fb}, we consider this contribution generic.  Thus, in gravity mediated theories (where $M$ is approximately the Planck scale or a bit below), the gaugino masses will typically end up a loop below the scalar masses.

In Planck- or string-scale mediation of supersymmetry breaking, one possibility for removing the dominant scalar mass operator in (\ref{gravmed}) is through sequestering, {\it i.e.}, separating the visible and hidden sectors in an extra dimension or `conformal throats' \cite{ Randall:1998uk, Luty:2001jh}.  There has been some debate about how generic such sequestering is (\cite{ Dine:2004dv, Kachru:2007xp, Berg:2010ha, Andrey:2010ry}).  We will make the assumption that sequestering is not generic.

We are led to a class of gravity mediation models in which the gaugino masses are roughly a loop factor below the scalars. A possibility we will not explore, is single-sector gauge mediation, where again gaugino masses tend to be much more than a loop-factor lighter than scalar masses.  

In addition, for ``A terms'' to be at the same scale as scalar masses, they would have to be generated by operators like 
\begin{equation}\label{gravmed}
\int d^2 \theta \frac{Y H_u Q U^c}{M},
\end{equation}
again requiring a gauge singlet in the hidden sector.  Thus our philosophy suggests A terms are small -- again, dominated by a one-loop suppressed contribution from anomaly mediation.  This will of course have an important impact on the Higgs mass predictions.

Of course the natural version of these models were ruled out when gauginos were not discovered in the 1 GeV range!  If these theories are realized in Nature, some kind  of ``pressure'' on the measure pushing towards higher supersymmetry scales is needed, which counteracts the tuning of the cosmological constant and the electroweak scale.  We will not attempt to address the notoriously ill-defined question of quantifying these pressures. We will simply assume that whatever the measure is, the likelihood of having a hidden sector that produces degenerate sfermions and gauginos is much smaller than that of a split spectrum with the obvious fine-tuning for electroweak symmetry breaking. We stress that with the spectra we are considering, the fine-tunings are at  $\sim 10^{-4} \to 10^{-6}$ level are obviously very severe from the perspective of naturalness, but are dwarfed by the $10^{-60} - 10^{-120}$ levels of fine-tuning for the cosmological constant or the usual $10^{-30}$ level tuning for the hierarchy problem.  

\vskip 0.15in {\noindent \bf Higgsinos, the $\mu$ term, and the Giudice-Masiero mechanism}
\vskip 0.1in

What about the Higgsinos? The $\mu$ term, $W \supset \mu H_u H_d$, breaks both the Peccei-Quinn (PQ) symmetry and potentially an R symmetry, and thus there can be trivial reasons why it is much smaller than the Planck scale.  The simplest operator that generates a $\mu$ term is the one suggested long ago by Giudice and Masiero \cite{Giudice:1988yz}:
\begin{equation}
\lambda \int d^4\theta H_u H_d
\label{GM}
\end{equation}
where $\lambda$ is an arbitrary coefficient.  In global (flat-space) supersymmetry, this operator represents a total derivative.  When including supergravity using the conformal compensator language \cite{Siegel:1978mj,Gates:1983nr,Kaplunovsky:1994fg}, one should multiply this operator by $\phi^\dagger/\phi$ (assuming conformal weights of fields correspond to their canonical dimensions).  The compensator $\phi \simeq 1 + \theta^2 m_3/2$, where $m_{3/2}$ is the gravitino mass, as long as the theory has no Planck scale vevs \cite{Bagger:1999rd}.  Integrating out the Higgsinos and some of the scalars will generate a gauge-mediated-like contribution to the gaugino masses at one loop.  The contribution will take gauginos off of the `anomaly-mediated trajectory' in a special way -- a right-magnitude, but wrong-sign contribution to gaugino masses \cite{Nelson:2002sa,Hsieh:2006ig}.  However, the threshold correction will be affected by squared soft masses for the scalars, and is suppressed when $m_{sc}^2 > m_{3/2}^2$.  

In addition, the operator itself appears highly tuned when seen from a different frame.  One can remove a chiral operator ${\cal O}$ from the K\"ahler potential ${\cal K}$ via the transformation
\begin{equation}
{\cal K} \rightarrow {\cal K} - ({\cal O} + {\cal O}^\dagger) \:\:\:\:\: {\rm and}  \:\:\:\:\:  W \rightarrow W e^{{\cal O}/M_{pl}^2} .
\end{equation}
For ${\cal O} = \lambda H_u H_d$, the term in (\ref{GM}) becomes terms in the superpotential, $(1 + \lambda(H_u H_d/ M_{pl}^2) + ... )(W_{hid} + W_0 + W_{vis})$, where $W_{hid}$ contains the fields involved in dynamical supersymmetry breaking and $W_0$ is the operator that generates a constant superpotential.  Thus a pure Giudice-Masiero (GM) term (\ref{GM}) is a result of a precise relationship between the coefficients of two operators, $H_u H_d W_{hid}$ and $H_u H_d W_{0}$ (up to $H^2/M_{pl}^2$ corrections).  This particular combination could result from a direct coupling of the curvature, whereas direct couplings to the constant superpotential and supersymmetry breaking sectors could be suppressed due to sequestering.  However, we are assuming that there is no sequestering, and  thus we do not have a predictive relationship between the effective $\mu$ and $B\mu$ terms.  For example, if only the $H_u H_d W_{0}$ operator existed, the threshold would be purely supersymmetric, and, in the limit of vanishing scalar soft masses, would keep gauginos on the anomaly-mediated trajectory (of the MSSM without Higgsinos).  

Having said this,  for the sake of simplifying parameter space, we will take the case of pure GM as the `central value' of the threshold correction in theory space.  Regardless of the details of the threshold, it is clear that it is trivial to generate $\mu\sim m_{3/2}$ in multiple ways.  

Of course since the $\mu$ term also breaks PQ symmetry, it is possible to imagine that the Higgsinos are lighter, near the same scale as the gauginos, as in the earliest models of split SUSY \cite{ArkaniHamed:2004fb}.

One can imagine suppressing these operators, as they explicitly break the PQ symmetry (under which $H_u H_d$ is multiplied by a phase).  For a pure GM term, approximate PQ symmetry implies $\lambda \ll 1$ and for $m_{scalar}\sim m_{3/2}$, this leads to a suppression of $\mu = \lambda m_{3/2}$ and $B\mu = \lambda m_{3/2}^2$ and thus $\mu \sim m_{scalar}/\tan \beta$.  In the limit where $\lambda\rightarrow 0$ -- or more generally Planck suppressed superpotential couplings between $H_u H_d$ and $W_0$ or the hidden sector are absent, $\tan\beta$ is large yet the Higgs mass requires low scalar masses, thus rendering the spectrum unviable.  In principle, the K\"ahler potential operator $X_{hid}^\dagger X_{hid} H_u H_d$ could generate $B\mu\sim m_{scalar}^2$ in which case $\mu$ would be generated by gaugino loops such that $\mu \sim (\alpha/4\pi) m_{gaugino}$.  However, we see no symmetry reason for this limit and do not explore this spectrum further (its phenomenology was explored in \cite{Hall:2011jd,Hall:2012zp}).

If $\mu^2 \gg m_{sc}^2$, then electroweak symmetry is only broken if the coefficient of the scalar bilinear $H_u H_d$ (the $B\mu$ term) approaches the limit $B\mu\rightarrow\mu^2$.  This is an interesting case, as $\tan{\beta} \rightarrow 1$ in this limit, a value which is disfavored in the `natural' MSSM both because the physical Higgs mass is too low, and because the top Yukawa coupling gains a Landau pole below the GUT scale.  Neither of these present a problem with heavier scalars.  However, this limit requires not only the tuning for electroweak symmetry breaking ($\lambda \rightarrow 1$), but also the magic of a pure GM mass.  There may not be a clear UV reason to naturally favor this point in parameter space, but it is at least interesting that it is allowed phenomenologically and should be considered.

\vskip 0.2in{\bf \noindent Spectrum and Unification}
\vskip 0.1in

\begin{figure*}[t]
\includegraphics[width=0.35\textwidth]{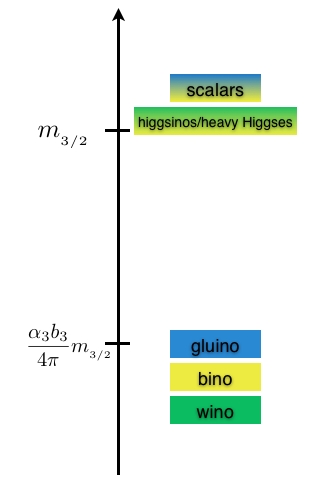}
\caption{A 'simply unnatural' spectrum.}
\label{fig:spectrum}
\end{figure*}

\begin{figure*}[b,h]
\includegraphics[width=0.85\textwidth]{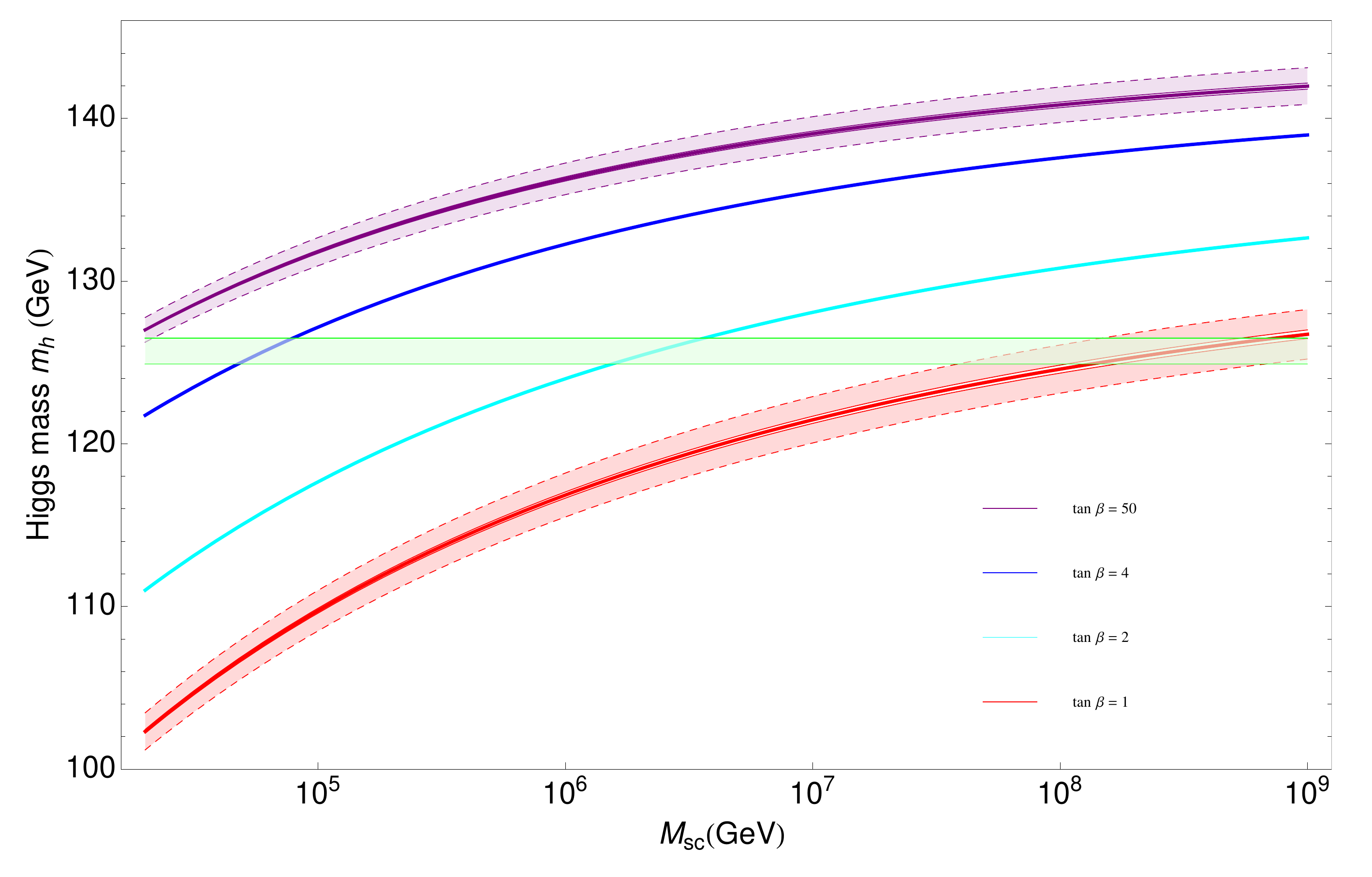}
\caption{Here we show the Higgs mass predicted as a function of the scalar masses and $\tan\beta$.  The bands at $\tan\beta=1$ and $50$ represent the theoretical uncertainty in the top mass and $\alpha_s$.  The gaugino spectrum is that predicted by the anomaly mediated contribution with the gravitino mass $m_{3/2} = 1000$ TeV, resulting in an approximate mass for the LSP wino of $\sim 2.7-3$ TeV (which is roughly the mass necessary for a the wino to have the correct cosmological thermal relic abundance to be all of dark matter \cite{Hisano:2006nn}).  The $\mu$ term is fixed to be equal to the scalar mass -- this threshold has a small but non-negligible effect on the Higgs mass relative to the conventional split supersymmetry spectrum \cite{ArkaniHamed:2004fb,Giudice:2004tc}.  The A-terms are small.}
\label{fig:higgsmass}
\end{figure*}

\begin{figure*}[t,h]
\includegraphics[width=0.75\textwidth]{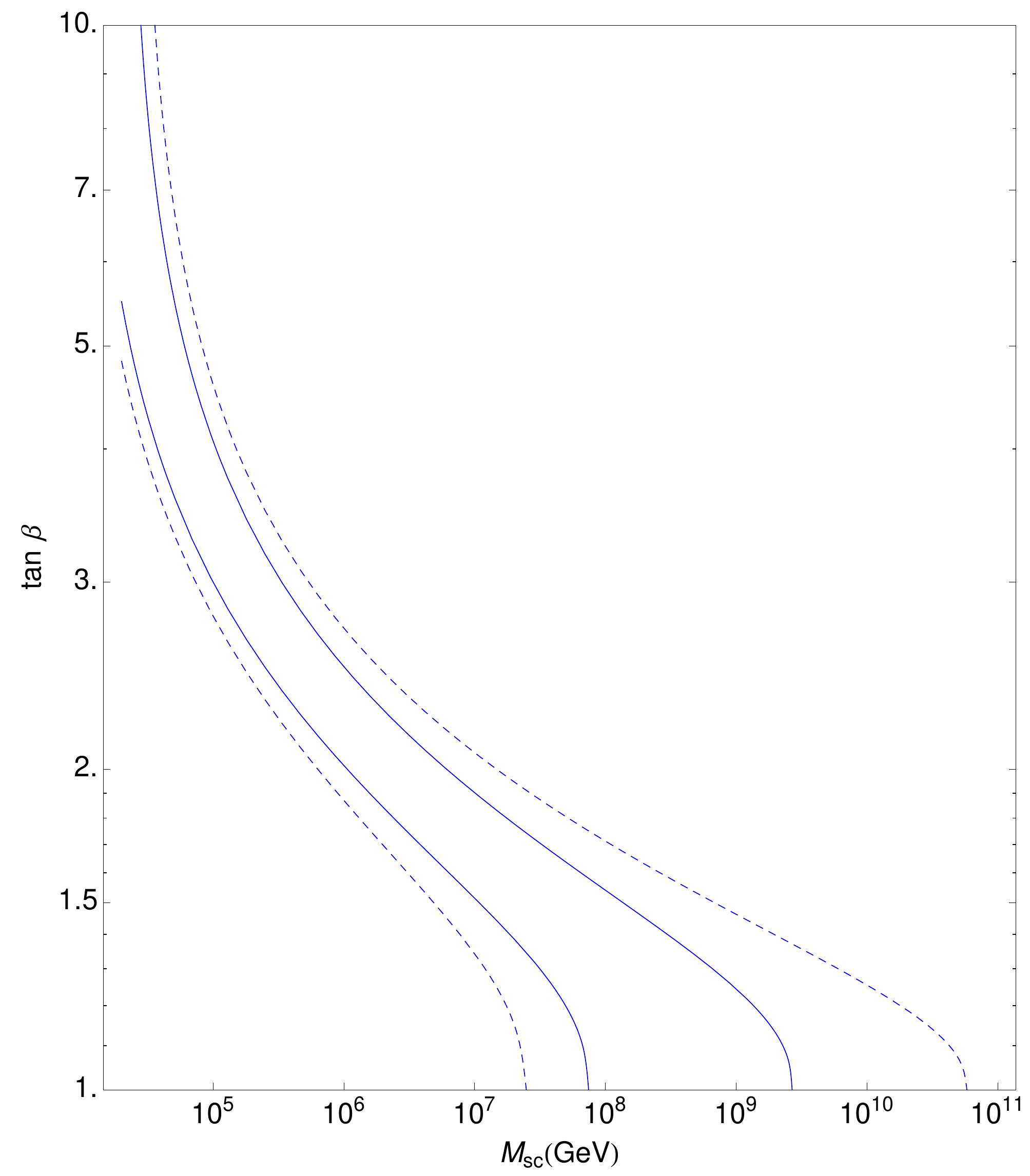}
\caption{The allowed parameter space in the $\tan\beta - M_{sc}$ plane for a Higgs mass of $125.7 \pm 0.8$ GeV, for $\mu = m_{sc}$. The solid blue lines delimit the $2 \sigma$ uncertainty. The dashed blue lines show the effect of the $1 \sigma$ uncertainty in the top mass, $m_t = 173.2 \pm 0.9$ GeV \cite{pdg}.  We take the gaugino spectrum predicted by AMSB (including the heavy Higgsino threshold) with the gravitino mass $m_{3/2} = 500$ TeV, resulting in a wino LSP at 2.6 TeV, and a gluino mass of 14.4 TeV.  However, the Higgs mass is highly insensitive to the gaugino spectrum, and a gravitino mass of 50 TeV yields essentially the same plot above.}
\label{fig:higgsmass2}
\end{figure*}

For our minimal model, we take scalar masses and Higgsinos to be roughly degenerate $m_{sc} \sim \mu \sim  m_{3/2}$ (Figure \ref{fig:spectrum}), and determine the overall scale favored by a Higgs mass of 125 GeV.  The result, shown in Figures \ref{fig:higgsmass} and \ref{fig:higgsmass2}, is somewhat different from that found in \cite{Giudice:2011cg}. The Higgsinos are not present in the effective theory beneath $m_{sc}$, and thus the positive contribution to the running of the Higgs quartic coupling from Higgsino/gaugino loops in the low-energy theory is absent. This means that $m_H \sim 125$ GeV is consistent with heavier scalars than in the split spectrum considered in \cite{Giudice:2011cg}. In particular, for moderate tan $\beta$, 
scalar masses in the $10^3 - 10^4$ TeV range are perfectly allowed, while with light Higgsinos  such heavy scalars are only possible if tan$\beta$ is tuned to be close to 1. Such heavy scalars naturally avoid all flavor problems, giving another impetus  to focus on a spectrum with $\mu\sim m_{3/2}$.

\begin{figure}
\includegraphics[width=0.85\textwidth]{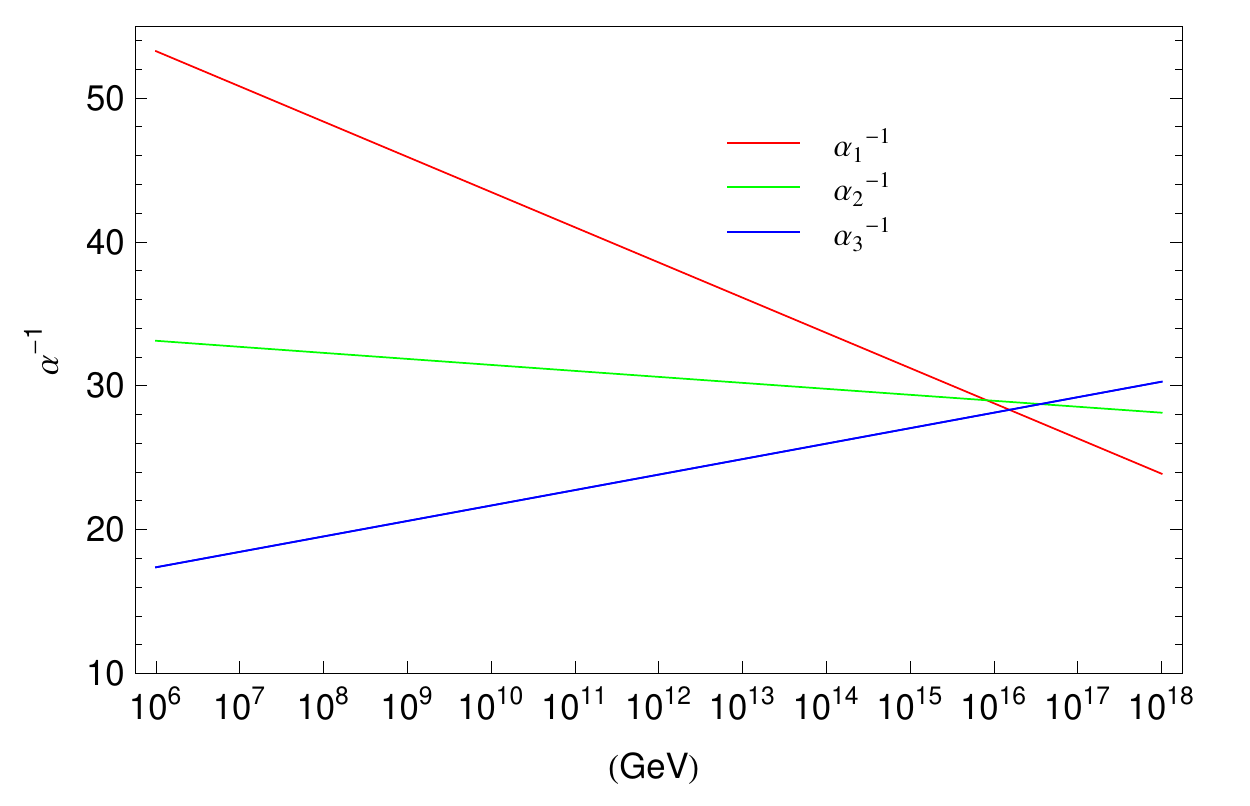}
\caption{Here we show the running of the gauge couplings with scalar masses and Higgsinos fixed at $10^3$ TeV.  Error bands on $\alpha_s$ are at the three-sigma level according to the Particle Data Group \cite{pdg}. We use $M_{gluino}$ = 14.4 TeV, $M_{wino}$ = 2.6 TeV, $M_{sc}$ = $\mu$ =  $10^3$ TeV and tan$\beta$ = 2.2 to generate this plot.}
\label{fig:unification}
\end{figure}

Because the Higgsinos have a significant impact on the differential running of gauge couplings, keeping $\mu$ heavy noticeably changes the unification prediction.  For example, we see in Figure \ref{fig:unification}, two-loop running predicts a smaller value of the strong coupling constant $\alpha(M_z)$ than what is generally found in the `natural' MSSM.  For example, with $\mu = 1000$ TeV, $\alpha(M_z)=.110$ for gluinos at 1.5 TeV and $\alpha(M_z)=.108$ for gluinos at 15 TeV (for $\mu=100$ TeV, $\alpha_s(M_z)=.115$ and $.113$ respectively).  Of course this prediction is affected by unknown threshold corrections at the GUT scale, but the values found here are a bit closer to the world average of $\alpha(M_z) = .1184(7)$ \cite{pdg}, and is very close to more recent determinations using LEP data \cite{Abbate:2010xh}.

\begin{figure}
\includegraphics[width=0.85\textwidth]{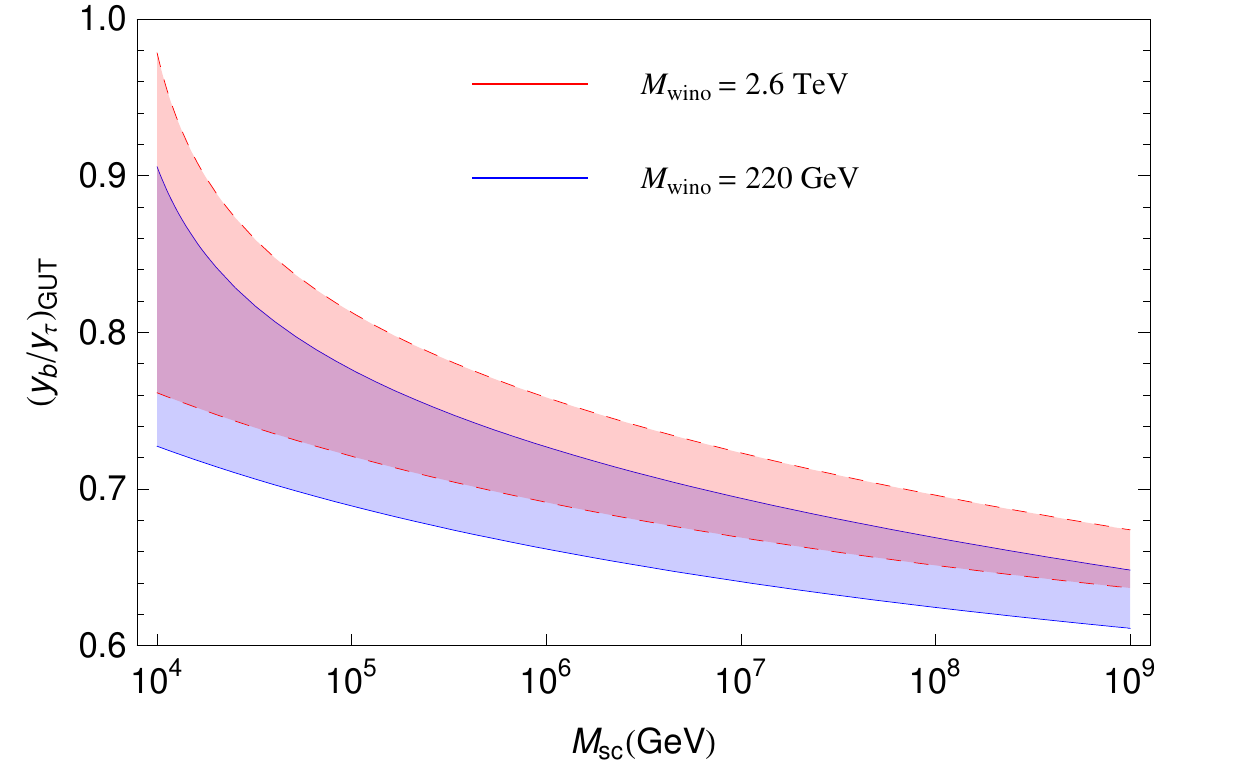}
\caption{This plot shows the Yukawa coupling ratio ($y_b/y_{\tau}$) evaluated at GUT scale as a function of the scalar mass. The gluino masses corresponding to the light and heavy wino masses are 1.5 TeV and 14.4 TeV, respectively. The bands correspond to tan$\beta$ = 1.2 and 50 from top to bottom.}
\label{fig:btau}
\end{figure}

Heavy scalars and Higgsinos do moderately better at $b$-$\tau$ unification than the natural MSSM (Figure \ref{fig:btau}), especially at low values of $\tan\beta$.  In addition, for small $\tan\beta$, the top Yukawa runs relatively strong at the GUT scale, and one would naturally expect significant threshold corrections.


In pure anomaly mediation, the gaugino masses are widely split, with the gluino roughly a factor of ten heavier than then wino.  This is due to the same accident as the near cancellation of the one-loop beta function of SU(2) in the MSSM.  With a pure GM term (ignoring soft masses), the Higgsino threshold increases the wino and bino masses such that the gluino/wino ratio is reduced to roughly a factor of six. An interesting limit occurs if the Higgses are mildly sequestered from $W_{hid}$ such that Planck-suppressed couplings to supersymmetry breaking are absent, but the $\mu$-term comes from $H_u H_d W_0$.  In such a limit, the threshold correction suppresses the wino mass, and in fact at leading order in $B\mu/\mu^2$ the wino mass vanishes!  Of course, without soft masses, electroweak symmetry breaking at a scale much smaller than $m_{3/2}$ would require $B\mu/\mu^2 \rightarrow 1$, in which case the wino retains $\sim 40\%$ of its standard MSSM value.  Without sequestering, however, soft masses generally reduce the threshold effect, and the operator $H_u H_d W_{hid}$ adds to the magnitude of the wino mass and thus reduces the large splitting.

\subsection{New Vector-Like States}

As with the $\mu$-term, $m_{3/2}$ is a natural mass scale for vector-like states.  Additional vector-like states, with big SUSY breaking, can further significantly modify change the anomaly-mediated spectrum of gauginos. 
To preserve gauge coupling unification, we assume that these states are in complete multiplets of $SU(5)$. In the simple limit that their masses come from a pure GM mass term, they invariably produce a squeezed spectrum among the MSSM gauginos \cite{natural-paper}.  As defined in \cite{natural-paper}, the effective number of messengers measures the size of the threshold correction compared to that of one canonical $5+{\bar 5}$ pair (in standard SU(5) language) with a pure GM mass and no additional scalar soft masses.

A heavy vector-like state whose mass comes only from a superpotential  ({\it i.e.}, supersymmetric) operator would, at leading order in $F/M$, decouple in such a way as to leave the anomaly-mediated relationships between beta-functions and gaugino masses intact.  In the case of a pure GM mass term, the effective $B\mu$ term for the scalar components of the new states have the opposite sign from the compensator as in the superpotential case, and therefore the threshold corrections to gauginos also have the opposite sign as that required to keep the spectrum 'anomaly mediated'.  For one to four sets of vector-like states,  this tends to suppress the splitting between the gluino and the wino (or lightest gaugino).  For example, with one vector-like state, the one-loop beta-function coefficients above the threshold for SU(3) and SU(2) are $b_3 = (b_3)_{\rm MSSM} +1 = -2$ and $b_2 = (b_2)_{\rm MSSM} +1= 2$ respectively.  Below the threshold, the coefficients become $(b_3)_{\rm MSSM}=-3$ and $(b_2)_{\rm MSSM}=1$, while the gaugino masses (at leading order) are proportional to $(b_3)_{\rm MSSM} + 2 = -1$ and $(b_2)_{\rm MSSM} +2 = 3$.  Accounting for the hierarchy in gauge couplings, this renders gluinos and winos roughly degenerate.  Generally, the gaugino-mass coefficients for $N$ messengers will be $(b_i)_{\rm MSSM} +2N$, where $i$ runs over the gauge groups.

\begin{figure}
\includegraphics[width=0.85\textwidth]{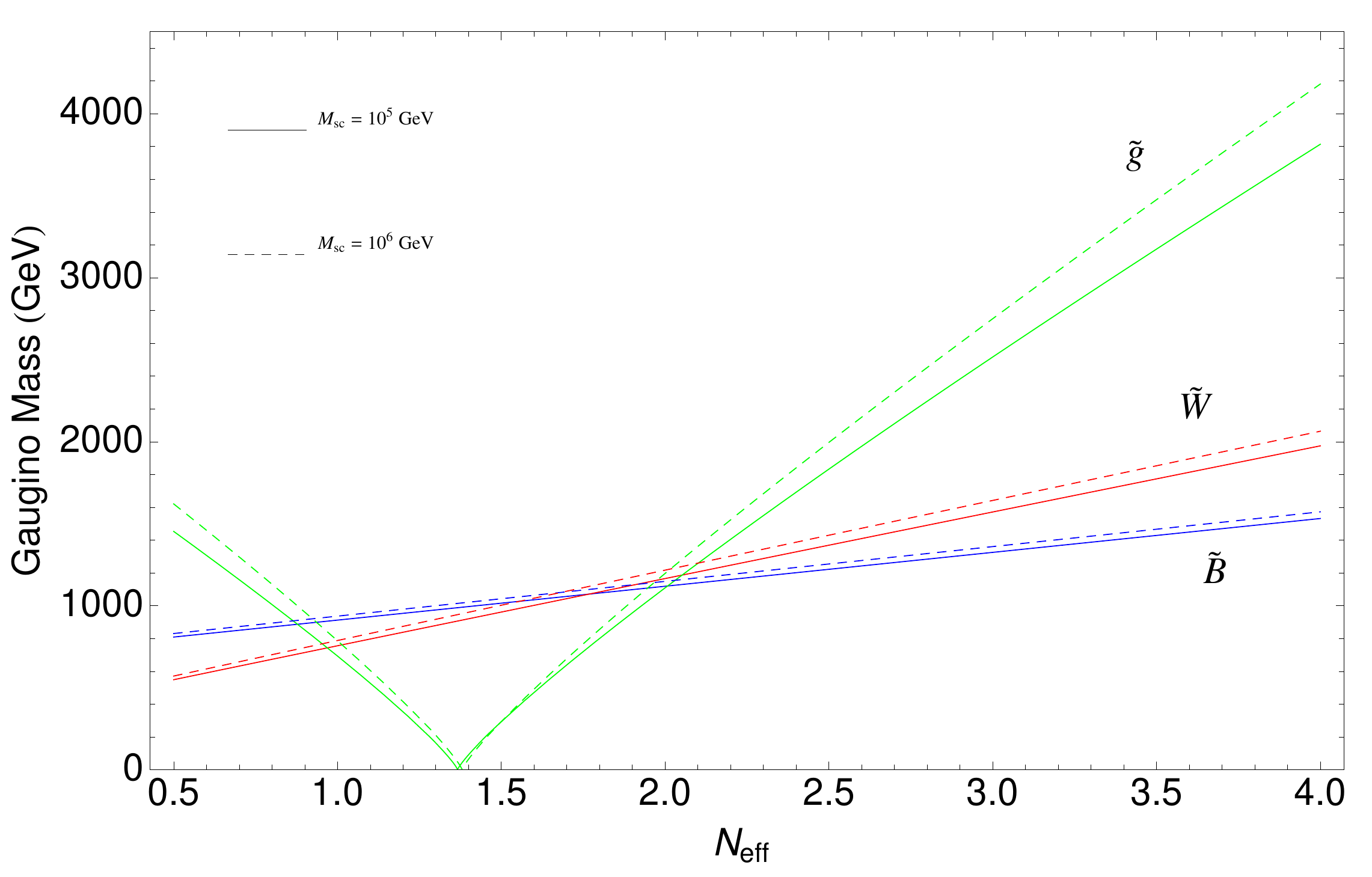}
\caption{This plot gives the gaugino spectrum as a function of $N_{eff}$ (defined in \cite{natural-paper}) at two loop plus threshold corrections.  The other parameters in the model are $m_{3/2} = 70$ TeV, $\tan{\beta} = 2.2$, and the coefficient for the GM term $\lambda=1.1$.  $M_{sc}$ is again the soft mass for all MSSM scalar superpartners and we set $\mu=M_{sc}$.}
\label{fig:Neff}
\end{figure}

More generally, soft masses for the scalar components of the vector-like state will suppress the threshold correction.  In the limit of soft masses much larger than the GM mass, the threshold correction goes to zero, and the resulting spectrum becomes proportional to $(b_i)_{\rm MSSM} + N$ -- only half the effect.  Thus, a more general parameterization of this threshold contribution is  $(b_i)_{\rm MSSM} +2N_{eff}$, where $N_{eff}$ (defined in \cite{natural-paper}) is $N$ in the GM limit, and $N/2$ in the limit of large scalar soft masses.  In Figure \ref{fig:Neff}, we plot the gaugino spectrum as a function of $N_{eff}$.  We see that the ratio of gluino mass to lightest gaugino is always smaller than the pure MSSM case.  A squeezed spectrum is of course more hopeful for the discovery of the gluino at collider for fixed LSP mass. 
\begin{figure}
\includegraphics[width=0.85\textwidth]{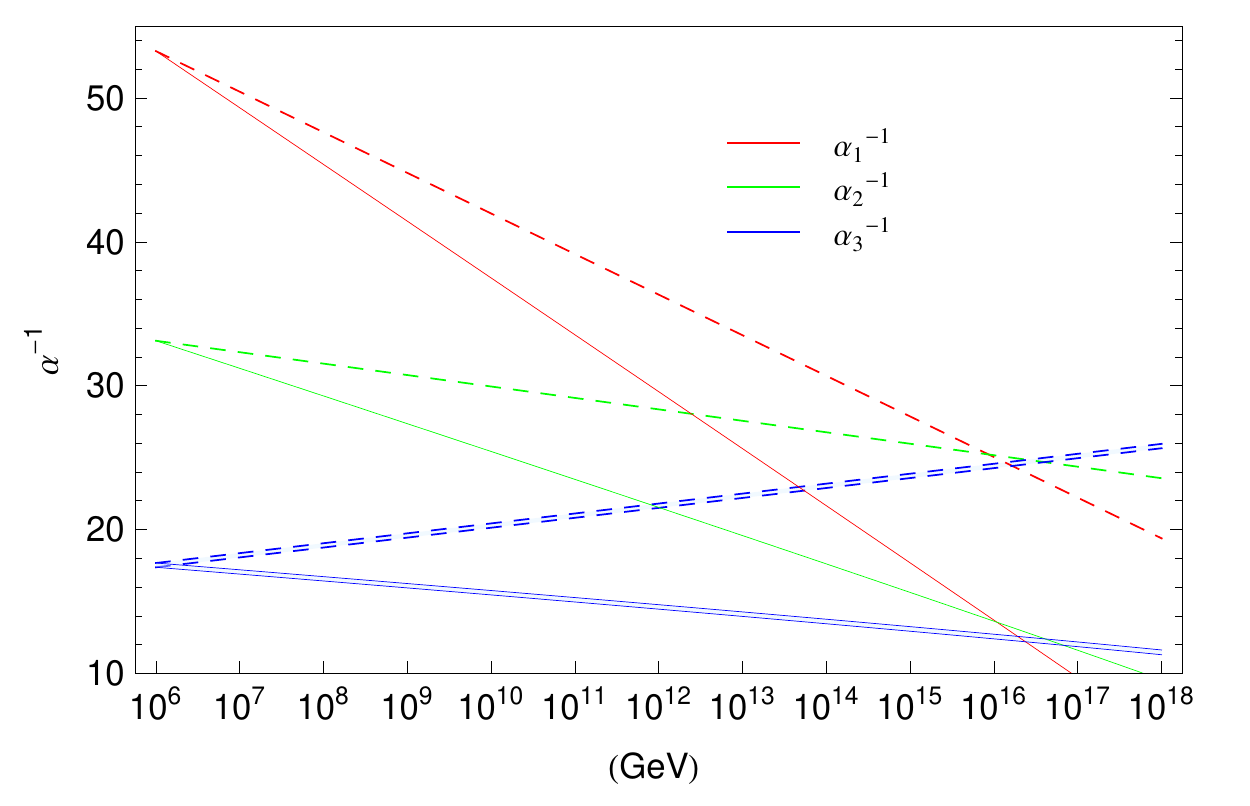}
\caption{Here we show the running gauge couplings in the case of $N=1$ vector-like state (dashed), and $N=4$ (solid). The scalar masses and Higgsinos are fixed at $10^3$ TeV.  Error bands on $\alpha_s$ are at the three-sigma level according to the Particle Data Group \cite{pdg}. We use $M_{gluino}$ = 14.4 TeV, $M_{wino}$ = 2.6 TeV, $M_{sc}$ = $\mu$ =  $10^3$ TeV and tan$\beta$ = 2.2 to generate this plot}
\label{fig:unif-N}
\end{figure}

The vector-like states have a slightly negative effect on unification, as shown in Figure \ref{fig:unif-N}.  Due to the two-loop running below the threshold.  For example, if we take the VLS scale to be $10^6$ GeV, then for $N=1$, the central value predicted for $\alpha_s$ is 0.1077, while for $N=4$ it is 0.1063.  In addition, $b-\tau$ unification is significantly worse.  However, if these new states are associated with a model of flavor, Yukawa coupling unification would depend on the full theory.

\subsection{Dark Matter}
One of the compelling motivations for new particles at the weak scale is the idea of WIMP dark matter. In models of the sort we are considering, where R-parity makes the LSP stable, we expect some thermal relic abundance regardless of whether the LSP comprises the majority of the dark matter. And since this is the lightest new particle in the spectrum, it is important to understand what mass it can have. 

To begin with, we can consider the conventional anomaly-mediated spectrum, with a wino dark matter candidate \cite{Moroi:1999zb}. In this case, to achieve the appropriate relic abundance, we require a mass of $\sim 2.7\-- 3 \tev$ \cite{Hisano:2006nn}.  With conventional anomaly mediation for the gaugino masses, this would make the gluinos inaccessible at the LHC. However, as we have already discussed, with the contributions of the Higgsinos and potentially new vector-like states, the spectrum is naturally squeezed. If it is quite squeezed, it is conceivable that the gluinos will be just at the edge of discovery, even with a thermal relic wino dark matter candidate. Since the direct detection cross section of a pure wino is extremely small \cite{Hisano:2004pv,Hill:2011be}, below $O(10^{-47} {\rm cm}^2)$, discovery via direct detection will be extremely difficult. 

However, a number of other options are also possible. With a wino LSP, it may simply be that the dark matter is dominantly composed of something else (e.g., axions). In such a case, the LSP can quite light (from the perspective of cosmological constraints), and almost any spectrum is open to us, including relatively light ($\sim \tev$ gluinos). Such a wino could be the dark matter if produced non-thermally (e.g., \cite{Gherghetta:1999sw,Moroi:1999zb,Ibe:2004tg}, or more recently \cite{Grajek:2008jb}). Indeed, in  the context of minimal split SUSY models,  it is reasonable to expect late-decaying moduli to dilute any thermal LSP abundance, with the dark matter being re-populated by modulus decays. This still favors dark matter lighter than the TeV scale to get the correct relic abundance, and can also pleasingly dilute the troublesome axion abundance down to acceptable levels, for axion decay constants almost as high as the GUT scale \cite{Kaplan:2006vm}.  If the bino is the LSP, we must rely upon late-time entropy production to dilute away an otherwise highly overabundant relic.

In each case, there remains the prospect for interesting collider signals. For a thermal relic, we must count on a squeezed spectrum, while non-thermal (or non-WIMP dark matter) cases are generally easier to find. Regardless, the appearance of signals at the LHC will possibly point to a non-standard thermal history.

\section{New Flavor Physics and Radiative Fermion Masses}

In our picture, the supersymmetric flavor problem must be solved in a trivial way, and not with ingenious model-building and gymnastics. Without any special structure to the scalar mass matrices, in particular with no mechanism enforcing scalar mass degeneracy, $K-\bar{K}$ mixing and $\epsilon_K$ demand that the first two generations squarks to be as heavy as $\sim$ 1000's of TeV. What about the third generation squarks? They can plausibly be comparable to the first two generations, or at most an order of magnitude lighter. 

To give a simple example for theories of flavor leading to the second possibility, consider models where the Yukawa hierarchy is explained by the Frogatt-Nielsen mechanism, with the light generations having different charges under anomalous $U(1)$ symmetries \cite{Ibanez:1992fy, Nelson:1997bt}.  The anomalous $U(1)$'s are Higgsed by the Green-Schwartz mechanism, and the gauge bosons are lighter than the UV cutoff (string scale), parametrically by a factor of $\sqrt{\alpha}$. Tree-level exchange of this $U(1)$ gauge boson can give SUSY breaking that dominates over Planck-suppressed soft masses.  This gives large, different masses to the first two generations, since they are charged under the $U(1)$, but not to the third generation. With an O(1) splitting between the first two generation scalars, these soft masses must be in the range of at least 1000's of TeV. Planck suppressed operators will put the third generation scalars in the range of 100's of TeV. Note that we can't imagine the third generation much lighter than a factor of $\sim 10$ compared to the first two generations. In RG scaling from high scales, 2-loop correction to the third generation soft masses from the first two generations give large negative soft masses that would lead to color breaking \cite{ArkaniHamed:1997ab}.
                                                                                                                                                            
Thus, if we want to trivially solve the flavor problem, the first two generation scalars should be in the range of 1000's of TeV, while the stops can be at most an order of magnitude lighter, in the range of hundreds of TeV. Note that the Higgs mass constrains the geometric mean of the left and right stop masses $m_{\tilde t} = \sqrt{m_{Q_3} m_{U_3}}$, and as we have seen, with  tan$\beta \sim O(1)$, we can have $m_{\tilde t} \sim 10^2 - 10^3$ TeV, so this is perfectly consistent with solving the flavor problem.                                                                     
It is also interesting that we can't much the scalars much {\it heavier}, without making the Higgs mass too big. Thus, the absence of flavor-violations pushes the scalar masses up, but getting the right Higgs mass doesn't allow these masses to get too large,                                                                           
saturating right around the 1000's of TeV scale. It is of course notable that this is just what is expected from the simplest picture of SUSY breaking we have been discussing.                                                                   

If we have scalars in just this range, with no special effort to suppress flavor-violations in the soft terms,  we might be sensitive to new flavor violations in future experiments.                                                                                                                                         
                                                                                                                                                            
\begin{figure}
\includegraphics[width=0.45\textwidth]{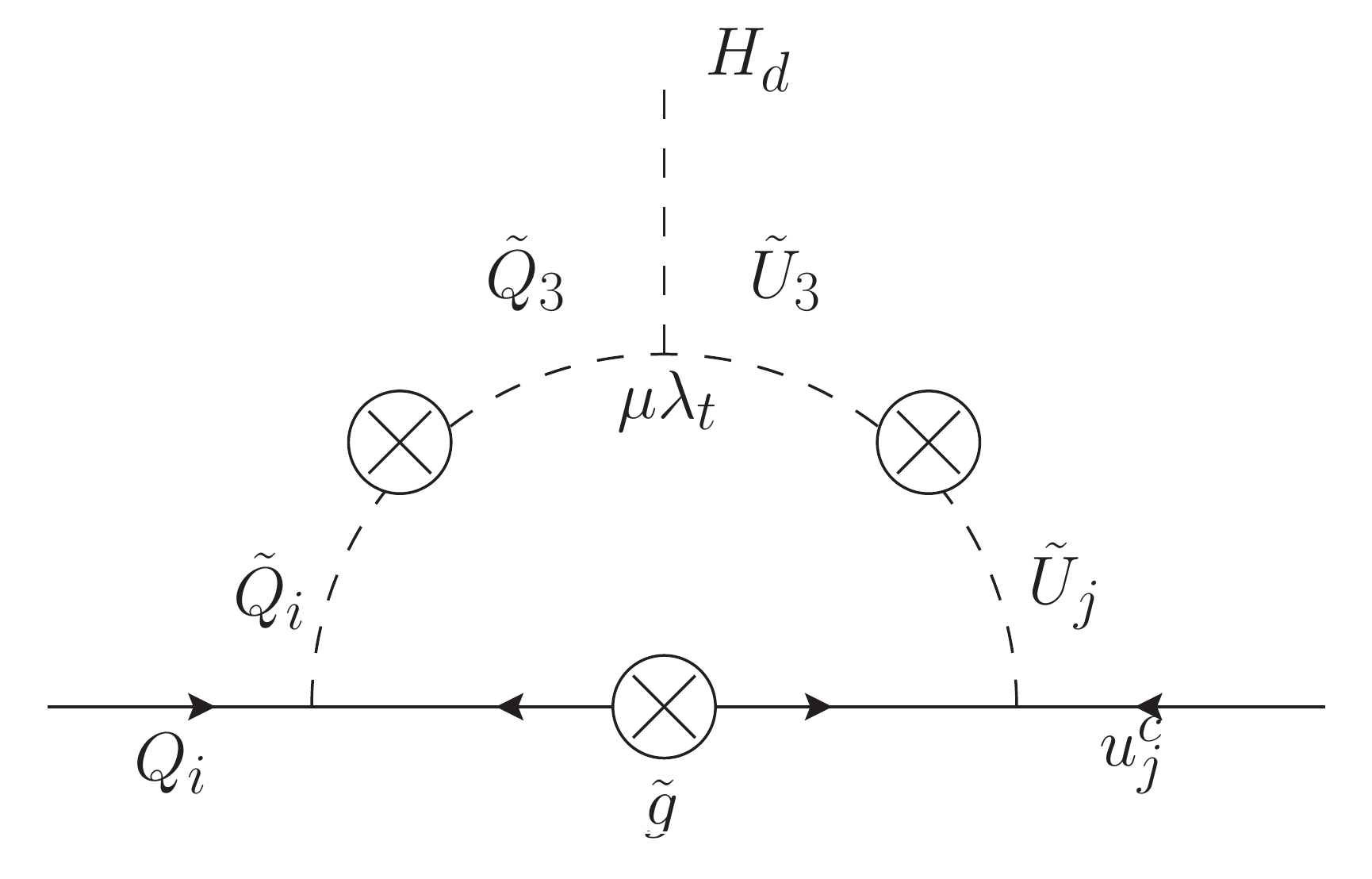}
\caption{}
\label{fig:oneloopfigure}
\end{figure}

There is another interesting observation about flavor, which provides an additional motivation for a ``split" spectrum with gauginos lighter than scalars. Let us suppose that there are indeed large flavor violations in the soft masses. The most obvious worry are huge FCNC's, but (ignoring detailed issues of the Higgs mass for the moment) one would imagine that  these could be decoupled  by making the scalars arbitrarily heavy. This is of course correct.  But in a theory with no splitting between scalars, Higgsinos and gauginos, there is a far greater difficulty with flavor that can not be decoupled by pushing up the scale of SUSY breaking: the large flavor-violations, in tandem with a large top Yukawa coupling and the breaking of R-symmetry by the $\mu$ term and gaugino masses, radiatively feeds an unacceptably large radiative contributions to the up-quark Yukawa couplings at one-loop. The diagram in Fig. \ref{fig:oneloopfigure} yields                                                                                                                     
which gives                                                                                                                                                 
\be                                                                                                                                                         
\delta \lambda_u^{ij} \sim \frac{\alpha_s}{4 \pi} \frac{m^2_{Q i 3}}{m_{sc}^2} \frac{m^2_{U j 3}}{m_{sc}^2}                                                                                                                                                
\frac{\lambda_t}{{\rm tan} \beta} \frac{\mu m_{\tilde{g}}}{m_{sc}^2}                                                                                           
\ee                                                                                                                                                         
With $m^2_{ij}/m_{sc}^2 \sim O(1)$, tan $\beta \sim O(1)$, and $\mu \sim m_{\tilde{g}} \sim m_{sc}$, this gives a correction to all up Yukawa couplings of order $\sim 10^{-2}$, vastly larger than observed.

It is interesting that with our minimally split spectrum, with $\mu \sim m_{sc}$ and $m_{\tilde g} \sim 10^{-2} m_{sc}$, this correction is roughly of the order of the up quark Yukawa coupling.  The up quark mass can plausibly arise from this ``SUSY slop".  Note that the analogous ``slop" can not be significant for the down and electron Yukawa couplings since the correction are $\propto \lambda_{b,\tau}$ tan $\beta$, and for the moderate tan$\beta$ we are forced to have,  the corrections are about $10^{-2}$ of the observed values.

More generally, supersymmetric theories with a split spectrum allow us to re-open the idea of a radiatively generated hierarchy for Yukawa couplings. The central challenge to building such theories of flavor is the following: the chiral symmetries protecting the generation of Yukawa couplings must obviously be broken, but then what forces the Yukawas to only be generated at higher loop orders \cite{Weinberg:1971nd, Georgi:1972hy}? Supersymmetry offers the perfect solution to this problem, since  the chiral symmetries can be broken in the K\"ahler potential, while holomorphy can prevent these breaking to be transmitted to Yukawa couplings in the superpotential. The chiral symmetry breaking is only transmitted to generate Yukawa couplings, radiatively, after SUSY breaking \cite{ArkaniHamed:1996zw}.   Unfortunately, it is extremely difficult to realize this idea in a simple way, with a natural supersymmetric spectrum; the flavor violations needed in the soft terms are large, and would lead to huge flavor-changing neutral currents. But in our new picture this is no longer the case:  Yukawa couplings are dimensionless and can be generated at any scale, while the FCNC's decouple as the scalars are made heavy.

As we have seen, with the minimal MSSM particle content, only the top Yukawa coupling is large enough to seed the other Yukawa couplings, and thus it is only possible to generate the up quark Yukawa coupling radiatively. Additional vector-like matter near $m_{sc}$ allows the possibility of new large Yukawa couplings and thus more radiatively generated Yukawas. For instance, with a single additional $(5 + \bar{5})$, $(D^c,L) + (D,L^c)$, we can have large mixing Yukawas of the form
\be
\lambda_d^i q_i  H_d D^c, \lambda_e^i e^c_i H_d L
\ee
as well as large scalar soft mass mixing
\be
m^2_{d i} \tilde{D^c}^* \tilde{d^{c i}},
m^2_{l i} \tilde{L}^* \tilde{l^i}
\ee
between the  $(D^c,L)$ and the ordinary $d^c_i,l_i$. Then, the analogous diagram to Fig \ref{fig:oneloopfigure} contributes to down-type quark and charged lepton Yukawa couplings, with the $\lambda_t \to \lambda_{d,e}^i$. For $\lambda_{d,e} \sim O(1)$, we can have a radiative origin for the down quark and electron Yukawa couplings.

Given that $m_{\tilde{g}} \sim \frac{\alpha}{4 \pi} m_{sc}$, these radiative corrections are parametrically 2-loop effects. With additional full vector-like generations, together with heavy-heavy and heavy-light Yukawa couplings to the Higgses, we can also get 1-loop corrections through diagrams of the form  of Fig \ref{fig:flavorloops}:
\begin{figure}
\includegraphics[width=0.45\textwidth]{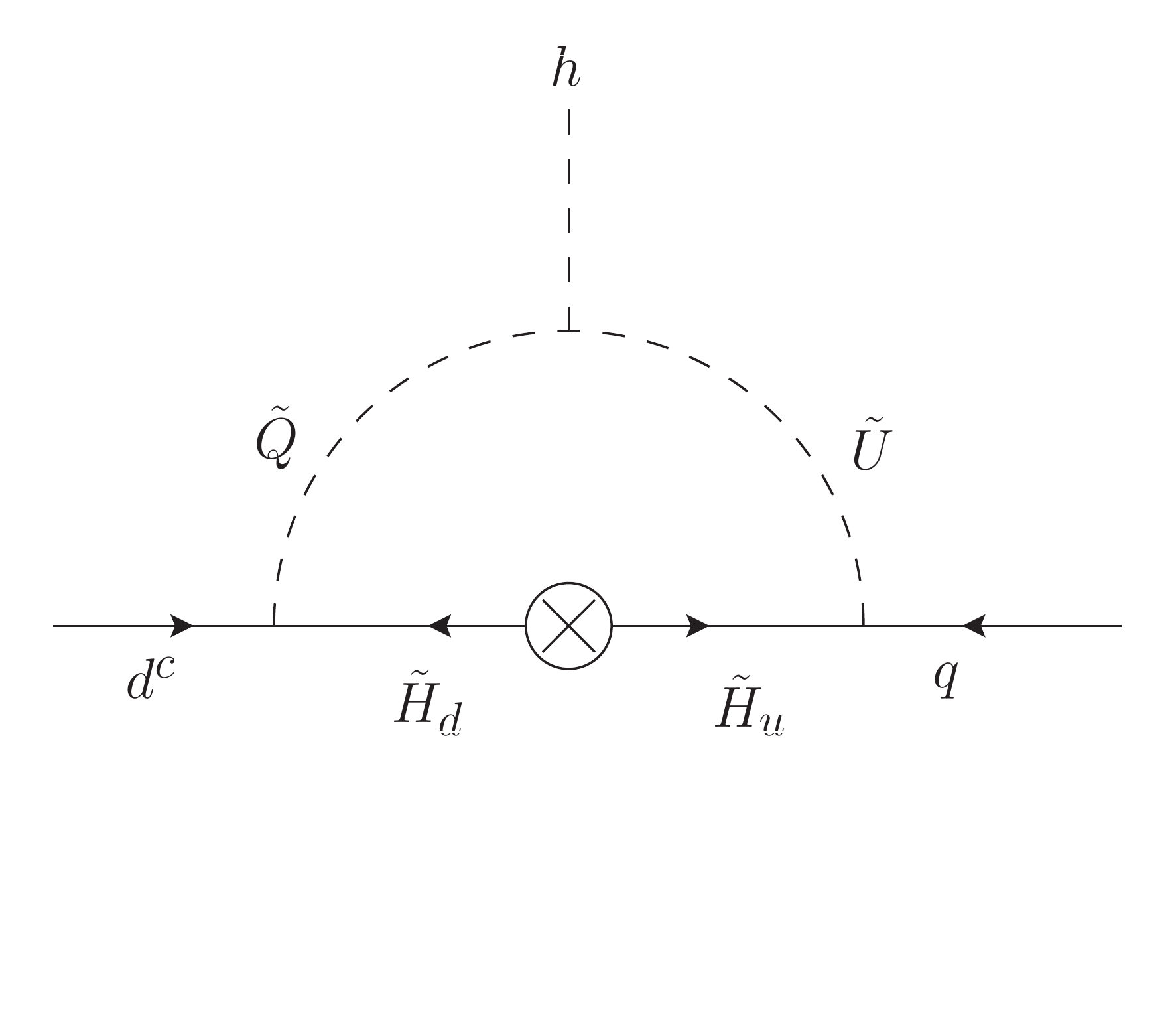}
\includegraphics[width=0.467\textwidth]{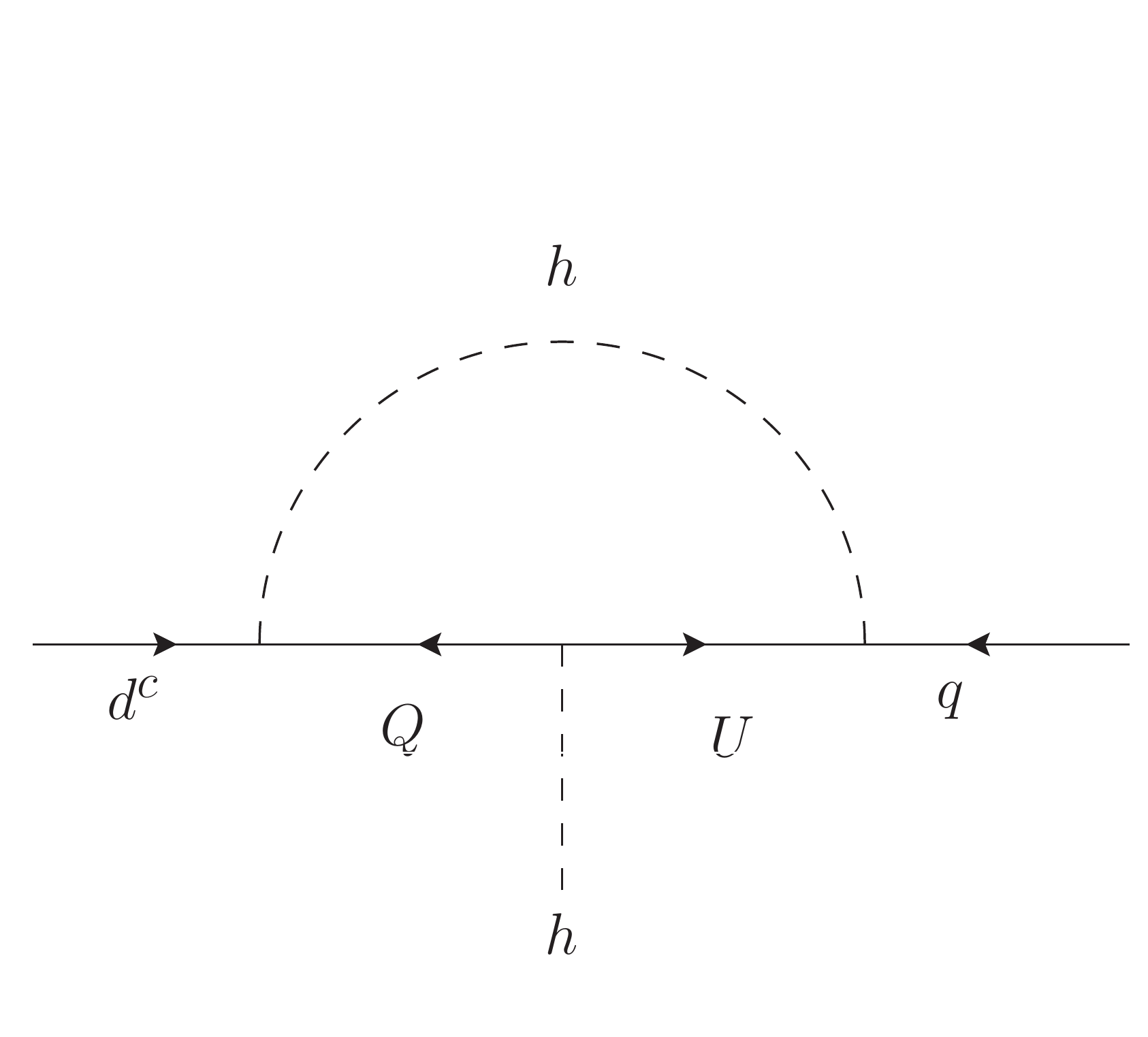}
\caption{}
\label{fig:flavorloops}
\end{figure}
This picture thus easily provides some basic ingredients for the construction of realistic theories where all the fermion masses arise radiatively off the top Yukawa coupling together with other $O(1)$ Yukawa couplings to heavy vector-like states. It would be interesting to attempt to build a complete theory of radiative fermion masses along these lines.

\section{Tests of Un-naturalness}

The theoretical developments leading to the development of the Standard Model were greatly aided by concrete experimental evidence for the presence of new physics at short distances not far removed from the experimentally accessible scales of the times.  This was most obvious for the weak interactions, which were encoded as dimension six operators suppressed by the Fermi scale. The presence of these operators, together with their $V-A$ structure, were strong clues pointing to the correct electroweak theory. The theoretical triumph of the Standard Model, has rewarded us with a renormalizable theory, with no direct evidence of higher dimension operators suppressed by nearby scales at all. Instead of having concrete clues to the structure of new physics through the observation of non-zero coefficients for  higher-dimension operators--say through a large correction to the S-parameter, a large rate for $\mu \to e \gamma$ or sizable electron EDMs-- the main guideline to extending the Standard Model for the past thirty years has been to explain ``zero": the {\it absence} of large quadratic divergent corrections to the Higgs mass, while seeing {\it no} observable effects in higher-dimension operators. 

The discovery of a  natural supersymmetric theory--as spectacular as it would be--would eventually leave us in a similar position: we would have another renormalizable theory, with no obvious indications for new physics needed till ultra-high energy scales. Amusingly enough, however, the situation is completely different in the un-natural theories we have been discussing in this paper.  The theory has two scales of new physics, for the gaugino masses $m_{1/2}$ and scalar masses $m_{sc}$. As we will see, if we can produce the gauginos, their decays can provide us with unique opportunities to measure or constrain higher-dimension operators suppressed by the scale $m_{sc}$.



Before turning to this discussion, let us first ask an even more basic question: what experimental signals can immediately {\it falsify} these simply un-natural theories? The existence of any new scalar state beyond the Higgs would immediately exclude simply un-natural models, since it would require an additional tuning, mechanism to stabilize its mass, or an elaborate family/Higgs symmetry structure. The only important caveat to this is that pNGBs could still be consistently present. However, if these are light without tuning, they can't be charged under the SM gauge groups, and they can only have higher-dimension couplings to SM fields suppressed by their decay constant. We would not expect to produce such states at colliders. 

A second light Higgs,  mixing and significantly altering the properties of the Higgs (e.g., \cite{Blum:2012kn,Blum:2012ii,Alves:2012ez}), would exclude these theories. This makes precision measurements of the Higgs especially important. Current bounds still allow for a sizable ($\gsim 2$) enhancement of the rate for $h\rightarrow \gamma \gamma$ \cite{ATLAS-CONF-2012-127,cms:2012gu,Espinosa:2012ir,Carmi:2012zd,Espinosa:2012im}, but with only fermions, it is a challenge to achieve rates above a factor of 1.5, and even then with some tension relative to precision electroweak observables \cite{Dawson:2012di,ArkaniHamed:2012kq,Kearney:2012zi}.  Large modifications of the Higgs couplings to the $W,Z$ require the existence of a new scalar. Higgs couplings to fermions can be modified by e.g., mixing with new vectorlike matter, as can the higgs width to $\gamma \gamma$. In all cases, a large $O(50\%)$ or larger boost to $\gamma \gamma$ signals, without an associated discovery of charged particles lighter than $\sim 150$ GeV, can conclusively exclude simply un-natural theories. 

In fact, within the framework we are discussing, with only the MSSM field content present, the leading interactions of the Higgs bosons to new, electroweak charged states are suppressed by $1/\mu$, and thus the corrections to Higgs properties are far too small to be seen. Thus, at least this minimal framework could be excluded by any convincing deviation of Higgs properties from SM expectations.



\subsection{Gaugino Decays and the Next Scale}
In a simply unnatural theory, the only new particles that we expect to see at the LHC are the gauginos. The impact of this on the decays of gauginos is profound, for a simple reason: {\em for $SU(3)\times SU(2) \times U(1)$ fermions that do not carry B or L, there are no renormalizable operators under which they can decay into each other and SM particles.} This simple point was emphasized by \cite{ArkaniHamed:2004fb}. As a consequence, the gaugino decays are necessarily suppressed by a new higher scale.

\begin{figure}
\includegraphics[width=0.45\textwidth]{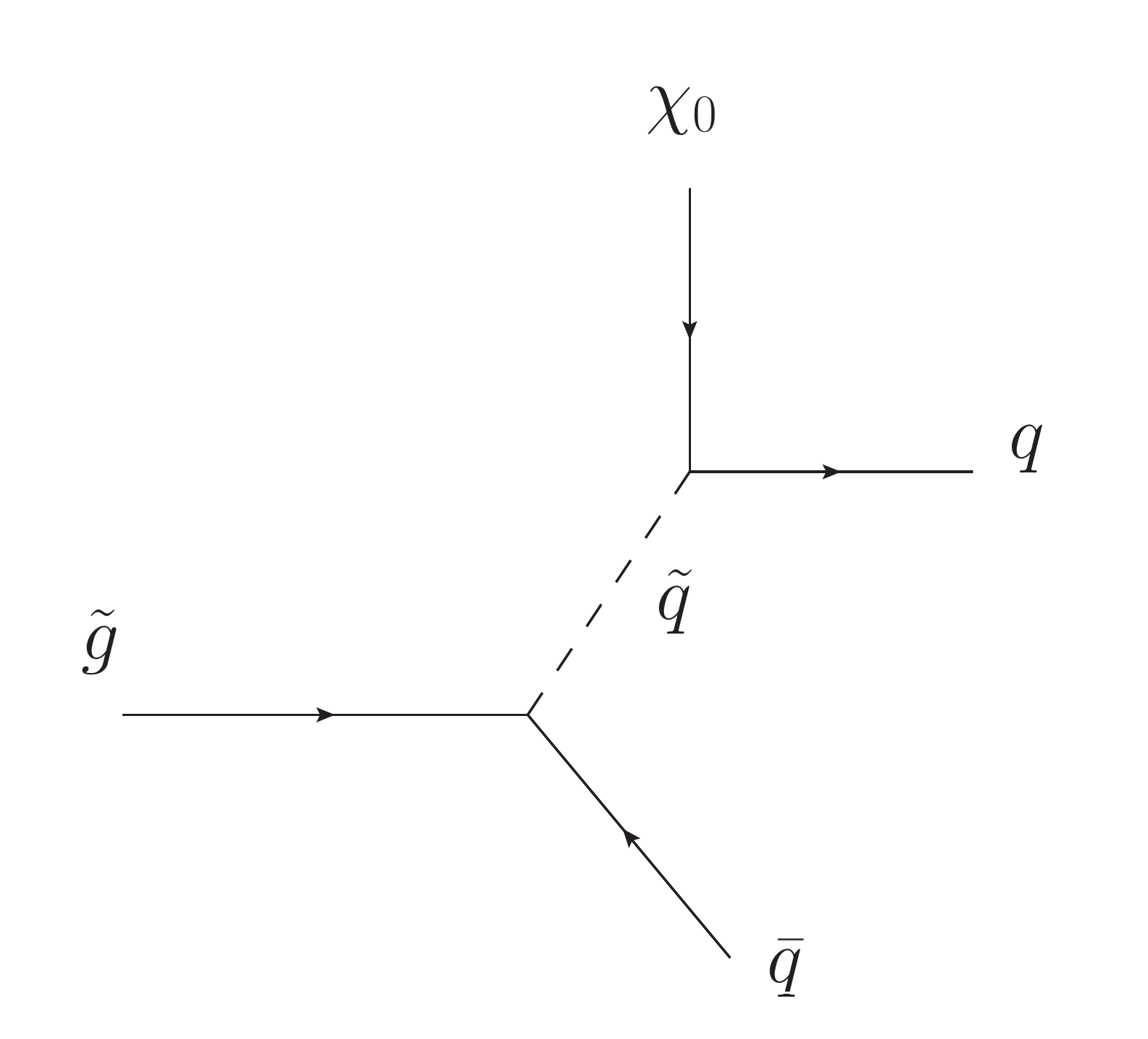}
\caption{}
\label{fig:gluinodecay}
\end{figure}

The scale in question varies depending on the particular process. Gluinos decay through the diagram in figure \ref{fig:gluinodecay}, which yields the dimension-6 operator
\be
\frac{g_3 \bar q \tilde g \bar \chi q}{m_{\tilde q}^2}.
\label{eq:dim6gluino}
\ee
The lifetime for such a decay can be quite long, with
\be
c \tau \approx 10^{-5} {\rm m} \left(\frac{m_{\tilde q}}{\rm PeV} \right)^4 \left(\frac{\rm TeV}{m_{\tilde g}} \right)^5~.
\ee

This leads to an interesting immediate observation: the fact that gluinos decay {\em at all} inside the detector will imply a scale within a few orders of magnitude of the gluino mass scale. Moreover, if the gluino decays promptly, without any displacement, we will already know that the scalar mass scale is at an energy scale $\lsim 100$ TeV, that is at least conceivably accessible to future accelerators. 

While this signal places an upper bound on the next mass scale, there are signals that can simultaneously place a quick lower bound. In particular, it is possible to imagine that large flavor violation in the scalar sector could produce clear flavor violation in the gluino decays (e.g., $\tilde g \rightarrow \bar t c$). If so, closing the loop generates sizable flavor violating four-fermi operators $\alpha_s^2 q^4/M_{\rm scalar}^2$. Even for CP conserving processes, 
constraints push this scale to \cite{Isidori:2010kg} $\sim 10^{3} \tev$ ($\sim 10^4 \tev$ if CP is violated). A combination of a lack of displaced vertices and large flavor violation in gluino decays could quite narrowly place the next scale of physics, without ever having observed a single particle close to the heavy scale. 

The quark line above can be closed to yield a chromomagnetic dipole operator as well
\be
\frac{g_3^3}{16 \pi^2}\frac{ m_{\tilde g}}{m_{\tilde q}^2} \log(m_{\tilde q}/m_{\tilde g}) \tilde g^i_j \sigma^{\mu\nu} \tilde b G_{i\mu\nu}^j.
\ee
Such an operator will produce dijet + MET signals, but because their rate is suppressed by a loop factor, they should be lost in the overall four jet + MET signals of the off-shell squark decay.

In contrast to gluinos, bino decay proceeds through a dimension-5 operator that arises from integrating out the Higgsino, namely
\be
\frac{g_2 g_1}{\mu} h^*_i \tilde W^i_j h^j \tilde B~.
\label{eq:bigdim5}
\ee
The suppression by only one power of the heavy scale suggests that these decays will be prompt.

Note that this operator generates a $\tilde W - \tilde B$ mixing term, which in general will correct the mass of the neutral wino relative to the charged wino by an amount $\sim \tilde m_W^4/(\mu^2 m_{\tilde W}) \sim 10^{-7} \gev ({\rm PeV}/\mu)^2 ({\rm TeV}/m_{\tilde W})$. This correction is negligible, even compared to the conventional loop-suppressed mass splitting $m_{\tilde W^\pm} - m_{\tilde W^0} \approx 150 \MeV$ \cite{Feng:1999fu}. The leading dimension 5 operator $\tilde W^i_j h^*_i h^l \tilde W^j_l$, while correcting the wino mass does not yield any mass splittings between the usual components. Thus, the mass splitting between charged and neutral winos is a clear test of heavy Higgsinos.

The operator of Eq. \ref{eq:bigdim5} leads to the decay $\tilde W_0 \rightarrow h \tilde B_0$. Note, however, that because there is no light {\em charged} Higgs, there can be no decay $\tilde W^\pm \rightarrow h^\pm \tilde B_0$ through this operator. Rather, the equivalent decay will arise from the resulting $\tilde W \-- \tilde B$ mixing, giving  $\tilde W^\pm \rightarrow W \tilde B$. While the $\tilde W^{\pm} \tilde W_0$ production cross-section is generally smaller than the Higgs production cross section, it is not far off from the associated production cross section (a few hundred $fb$ at $m_{\tilde W} \sim 200 \gev$ \cite{Lisanti:2011cj}. Consequently, in these models, there are new avenues for $Wh$ production that might be searched for.

Note that this dimension 5 operator does {\it not} give $\tilde W_0 \rightarrow \tilde B_0 Z$. This decay can arise from a dimension 6 operator, integrating out the Higgsinos at tree-level
\be
\frac{g_1 g_2}{\mu^2} h^\dagger D_\mu h \tilde W \bar{\sigma}^\mu \tilde B
\ee
and it can also as a dimension 5 operator integrating out the Higgsinos at 1-loop, generating a dipole operator
\be
\frac{q g_2 g_1}{16 \pi^2\mu} \tilde W^i_j \sigma^{\mu\nu}\tilde b g_2 F_{i\mu\nu}^j.
\label{eq:smalldim5}
\ee
\begin{figure}
\includegraphics[width=0.45\textwidth]{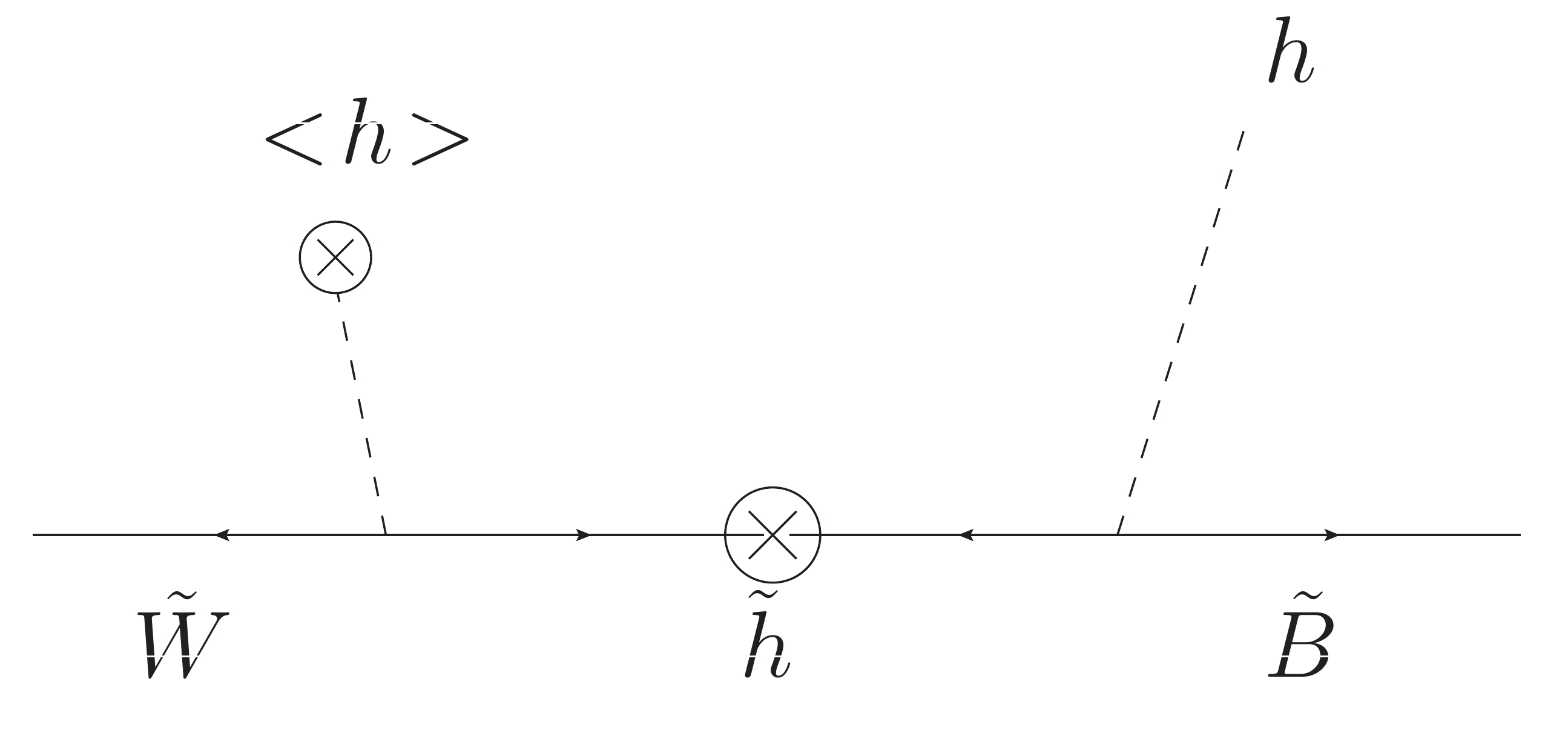}
\includegraphics[width=0.45\textwidth]{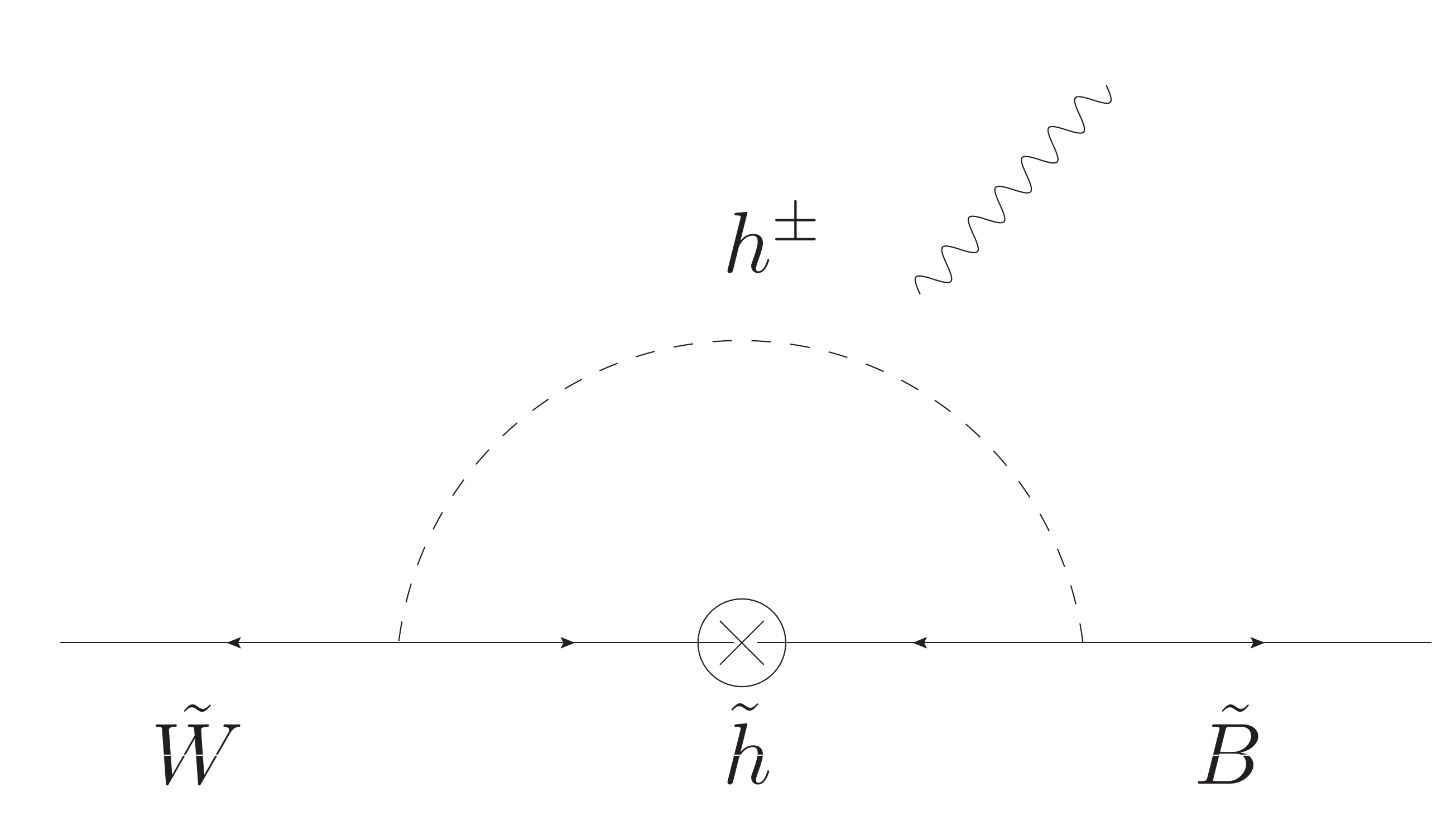}
\caption{}
\label{fig:dim5}
\end{figure}
\begin{figure}
\includegraphics[width=0.45\textwidth]{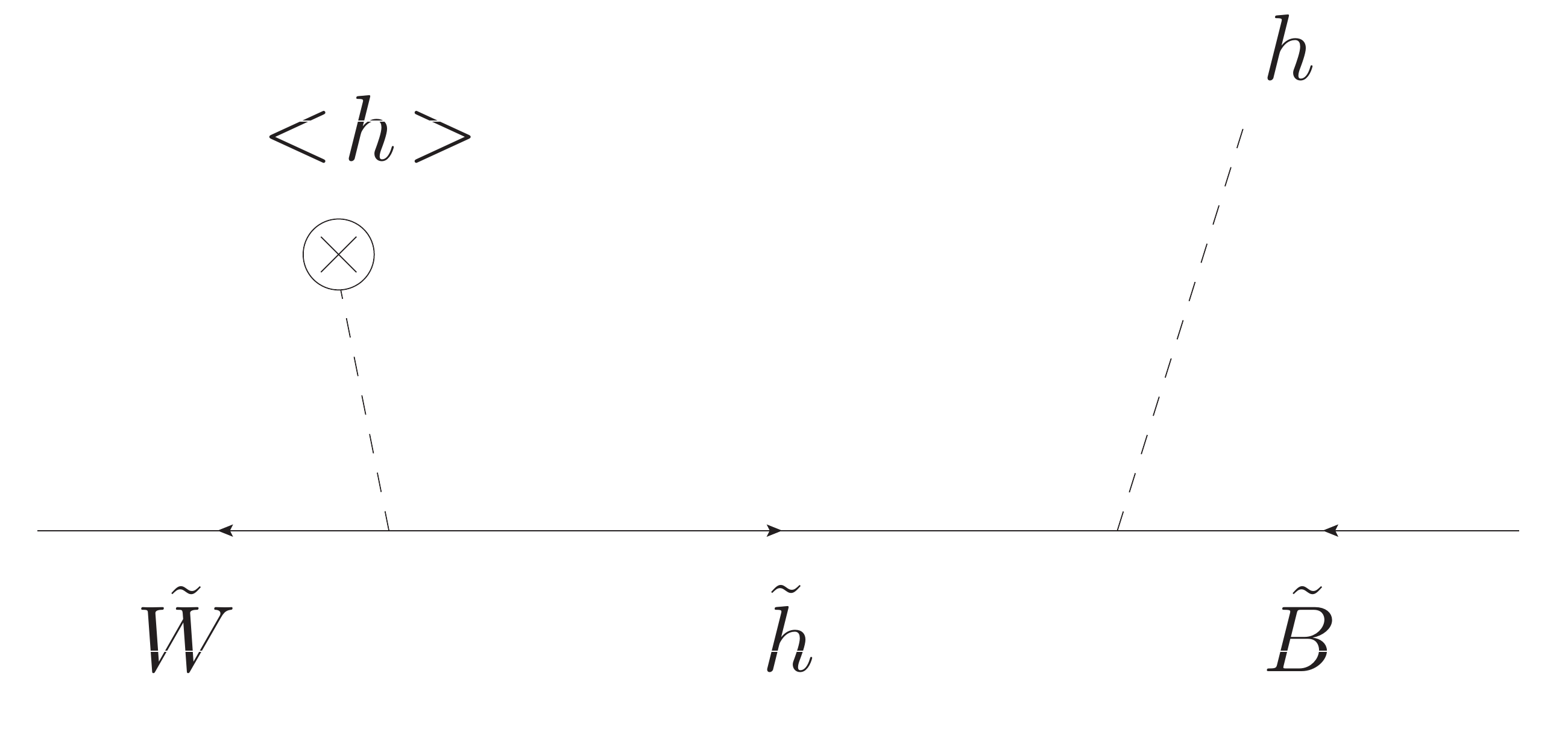}
\caption{}
\label{fig:dim6}
\end{figure}
In either case, with heavy Higgsinos  $\tilde W_0 \rightarrow \tilde B Z$ is expected to a rare decay. If the Higgsinos are heavier than $\sim 10$ TeV, the radiative dimension 5 operator dominates the amplitude and we have a branching ratio for this decay $\sim (\alpha/(4 \pi))^2 (m_{\tilde W}/m_Z)^2$. For heavy enough winos this could be observable. Note that the dimension six operator can only contribute to $\tilde W_0 \to \tilde B Z$ but not to $\tilde W_0 \to \tilde B \gamma$, while the dipole operator gives both. The pure dipole predicts a ratio of the photon to $Z$ final states of just $\sin^2\theta_W/\cos^2\theta_W \sim 1/3$. A measurement of this could establish the dipole operator as the source of the wino decay, and would show that the Higgsinos are heavy enough for the dimension six operator to be negligible. Alternately, a deviation from this ratio would tell us that the Higgsino is heavy, but lighter than $\sim 10$ TeV. 

Having enumerated the decay possibilities, let us now consider the signatures of gluino production and decay at the LHC. Let us assume a non-squeezed spectrum with $m_{\tilde g} > m_{\tilde W} > m_{\tilde B}$. This offers the possibility of spectacular processes. If the stops are the lightest colored scalars, we have the signal of $t^+t^-t^+t^-\tilde B \tilde B$ final states, which yields four tops + MET, where the stops are potentially produced from displaced vertices (if the scalar scale is high enough, or the spectrum is adequately squeezed). More striking is if the decay proceeds $\tilde g \rightarrow t^+t^- \tilde W^0$, with $\tilde W^0 \rightarrow \tilde B h$. In such a case we could find final states with 8 $b$'s, four $W^\pm$ and significant MET (and again, possibly displaced vertices). Such a process would have effectively zero background, making the only question for this scenario whether gluinos are produced at all. At 14 TeV and 300 $\rm fb^{-1}$,  we estimate approximately 5 events for $\sim$ 2.5 TeV gluinos (or 3 TeV gluinos for ten times that data). In some cases, the decay $\tilde g \rightarrow \bar t b \tilde W^+$ will occur, followed by $\tilde W^+ \rightarrow W^+ \tilde B^0$. Note that this final state is very similar (topologically) to the direct decay $\tilde g \rightarrow \bar t t \tilde B^0$. 

Let us consider the possibility that the bottom of the spectrum is reversed and wino is the LSP. Essentially all the decays should proceed via Higgs emission (if kinematically available). 
I.e., the decay $\tilde g \rightarrow \bar t t  \tilde B^0$ will be followed by $\tilde B^0 \rightarrow \tilde W h$. In contrast, direct decays to charged winos will proceed $\tilde g \rightarrow \bar t b \tilde W^-$, with the chargino proceeding to decay into $\tilde W^0$ producing a disappearing track.

Thus, for the $m_{\tilde W} > m_{\tilde B}$ case, the final states are \, $4 t$+ MET, as well $2t 2b 2W$+ MET, and $4 t 2 h$ + MET. For the $m_{\tilde B} > m_{\tilde W}$ case the final states are $4 t + $MET, $4t 2h$+MET and $2t 2b$+MET. It is clear from this list that distinguishing these cases will be nontrivial. However, the $W$ from the chargino decay should be distinguishable from one that came from top decay, while the direct decay to $b$ should produce a spectrum of $b$ quarks which are in principle distinct from those from top decay. And, of course, the presence of the classic disappearing track signature, once seen, would be a clear sign of the wino LSP.

\subsection{Gluino Decays and Stop Naturalness}
One of the key features of an unnatural theory is that the LR soft masses should be negligible. Even with large $A$ and $\mu$, these terms are also proportional to the Higgs vev, and are thus naturally $\sim 10^4$ times smaller than the soft mass-squared terms. This impacts gluino decays in an interesting way. 

In more detail, the gluino decay operators are
\be
\frac{g_2}{\Lambda^2_{t_l}}\tilde g b_L \bar t_L \tilde W^{-}
 \hskip 0.2 in \frac{g_2}{\Lambda^2_{t_l}}\tilde g t_L \bar t_L \tilde W^{0}
 \hskip 0.2 in \frac{g_1}{\Lambda^2_{t_l}}\tilde g t_L \bar t_L \tilde B^{0}
\label{eq:decayops}
\ee
\be
\nonumber \frac{g_1}{\Lambda^2_{t_l}}\tilde g b_L \bar b_L \tilde B^{0}
 \hskip 0.2 in \frac{g_1}{\Lambda^2_{t_r}}\tilde g t_R \bar t_R \tilde B^{0}
 \hskip 0.2 in \frac{g_1}{\Lambda^2_{b_r}}\tilde g b_R \bar b_R \tilde B^{0}
\ee

Where $\Lambda_{t_l}^{-2} = \sum g_3 U^L_{i3} \tilde m_{l ,i}^{-2}$, $\Lambda_{t_r}^{-2} =g_3  \sum U^{tr}_{i3} \tilde m_{tr,i}^{-2}$, and $\Lambda_{b_r}^{-2} =g_3  \sum U^{br}_{i3} \tilde m_{br,i}^{-2}$ are the weighted mass-squared where the matrix $U$ transforms between the flavor basis and the mass basis.

The key observation here is that we have five distinct decay modes into heavy flavor, $\tilde g \rightarrow \bar t t \tilde W^0$, $\tilde g \rightarrow \bar b b \tilde W^0$, $\tilde g \rightarrow \bar t b \tilde W^+$, $\tilde g \rightarrow \bar t t \tilde B^0$ and $\tilde g \rightarrow \bar b b \tilde B^0$. In contrast, we have only three distinct mass scales in the problem, $\Lambda_{t_l}, \Lambda_{t_r}$ and $\Lambda_{b_r}$. Thus, the decay of gluinos into heavy flavor is a highly overconstrained system in the unnatural limit, while for natural theories, cross terms introduce additional parameters into the theory. The heavy flavor branching ratios can easily falsify the unnatural scenario. Alternately, if they are consistent with small $A$ terms, this would place additional fine-tuning strain on the MSSM to accommodate  the Higgs mass, though of course, we cannot discount a cancellation that reduces sensitivity to these cross terms.

This discussion does not account for  top polarization measurements. Should $t_l$ and $t_r$ be distinguishable, this would introduce yet another quantity into an already overconstrained system. If that, too, could be understood with only the three mass scales, it would provide strong evidence of a simply unnatural theory. Regardless, it would clearly show that scalar masses are not significantly corrected by electroweak symmetry breaking. 

\section{Conclusions}

The expectation of a natural resolution to the hierarchy problem has always been the best reason to expect new physics at the TeV scale, accessible to the LHC. Naturalness demands new colored states lighter than a few hundred GeV, needed to stabilize the top loop corrections to the Higgs mass. These colored particles must be accessible at the TeV scale. Dark matter is another reason to expect new particles in the vicinity of the weak scale, but the WIMP ``miracle" is not particularly sharp and allows for masses and cross-sections that can vary over several orders of magnitude. Indeed, if we take the simplest picture for dark matter--new electroweak doublets or triplets, annihilating through the $W$ and $Z$, the needed masses are at 1 or 3 TeV, well out of range of direct production at the LHC. It is only naturalness that forces {\it colored} particles to be light, with the expectation that they should be copiously produced at the LHC.

On the other hand naturalness has been under indirect pressure from the earliest days of BSM model-building, and the pressure has been continuously intensifying on a number of fronts in the intervening years. The LHC is now exploring the territory where natural new physics should have shown up. No new physics has yet been seen, and while it is far from the time to abandon the idea of a completely natural theory for electroweak symmetry breaking, the confluence of indirect and direct evidence pointing against naturalness is becoming more compelling. But the Higgs mass $m_H \sim 125$ GeV, is within a stones throw of its expected value in supersymmetric theories, and the compelling aspects of low-energy SUSY--precision gauge-coupling unification and dark matter--remain unaltered. 

The simplest picture resolving the tensions given by this state of affairs is the minimally split SUSY model we have discussed in this paper. These models can be easily killed experimentally;  for instance if any of the hints for large enhancements of $h \to \gamma \gamma$ without enhancements to $h \to ZZ, WW$ are solidified. As for positive signals, indirect evidence for the heavy scalars can arise since since they are just in the range where they may give rise to interesting levels of FCNC's.  

The direct LHC probes of these models walks on a knife's edge of excitement. Obviously if the new fermions are too heavy to be produced we have nothing. But if the gauginos are directly produced, not only do we see new particles, but since their decays can only proceed through higher-dimension operators, we get a number of handles on the presence of next high scale between $10$ to  $1000$ TeV, ranging from displacement or flavor violation in gluino decays, to rare decay modes for the wino/bino. This would be enormously exciting, not only providing dramatic evidence for fine-tuning at the electroweak scale, but giving an indication of next thresholds that are not out of reasonable reach for new accelerators in this century.

As has long been appreciated and repeatedly pointed out, the dark matter motivation does not guarantee that the gauginos will be accessible to the LHC; the LSP could be a 3 TeV wino giving the correct relic abundance. But it is also perfectly possible that they are light enough to be produced. Fortunately, the final states from gluino decays are so spectacular that only a handful  need to be produced to confirm discovery. If Nature has indeed chosen the path of un-natural simplicity, we will have to hope that she will be kind enough to let us discover this by giving us a spectrum with electroweak-inos lighter than $\sim 300$ GeV or gluinos lighter than $\sim 3$ TeV.

\acknowledgements
N.~A.-H.~ is supported by the Department of Energy, grant number DE-FG02-91ER40654, A.~G.~, D.~E.~K.~, and T.~Z.~ by the National Science Foundation grant number 0244990, and N.~W.~ by the National Science Foundation grant number 0947827.

\bibliography{unnaturalbib}

\end{document}